\documentclass[11pt]{article}
\usepackage{hyperref}
\usepackage{color}
\usepackage{amsmath}
\pdfoutput=1
\begin{document}
\title{Gravity-driven thin-film flow with negatively buoyant particles}
\author{A. Mavromoustaki, P. David, S. Hill, P. Latterman\\ W. Rosenthal, M. Mata, A. L. Bertozzi  \\
\\\vspace{6pt} Department of Mathematics \\ University of California Los Angeles}
\maketitle
%% The abstract (in this file, and that submitted as text to arXiv) should
%include the exact phrase
%% "fluid dynamics video" or "fluid dynamics videos"
\begin{abstract}
This arXiv article describes the fluid dynamics video on `Gravity-driven thin-film flow with negatively buoyant particles', presented at the 64th Annual Meeting of the APS Division of Fluid Dynamics in Baltimore, MD in November 2011. The video shows three different experiments where a thin film of silicone oil laden with particles, is allowed to flow down an incline under the action of gravity. The videos were recorded at the UCLA Applied Math Laboratory.

\end{abstract}
% main text
\section{Video technical details}
The video (Entry \# V011, APS 64$^{\text{th}}$ Annual DFD Meeting 2011) demonstrates the patterns exhibited by gravity-driven, particle-laden thin films flowing down
a solid substrate. The results from three experiments are shown in the video. A finite volume of fluid mixed
with particles is allowed to flow down the solid plane;
visualization is achieved from the front view and the videos have
been recorded with a digital SLR camera (Canon EOS Rebel T2i). The substrate angle of inclination is fixed at $\theta=35^{o}$ while the
dimensionless particle concentration, $\phi$ (scaled
on a characteristic concentration associated with maximum packing) increases from left (blue
particles, $\phi=0.25$) to middle (red particles, $\phi=0.35$) to
right (yellow particles, $\phi=0.50$). The runtime associated with
the individual videos has been fast-forwarded in order to
demonstrate concurrently the onset of flow instabilities as well as the development of fingering
patterns. From left to right, the runtime speed has been increased
by 2, 4 and 20 times, respectively.

The choice of parameters allows the presentation of three, distinct
regimes. In this video, we  attribute solely the difference in
interfacial characteristics between the three regimes, to variations in
particle concentration. For small concentrations (left video, blue
particles), the particles settle rapidly allowing the clear fluid to
flow over them which results in fingering. For relatively large concentrations
(right video, yellow particles), the particles aggregate at the
contact line, forming a particle-rich ridge which appears to suppress the fingering. For intermediate concentrations (middle
video, red particles), we observe a well-mixed regime characterized
by finger formation.\\ \ \\ 

\noindent \textbf{References}\ \\ \

\noindent 1. T. Ward, C. Wey, R. Glidden, A. E. Hosoi, and A. L. Bertozzi. Experimental study of gravitation effects in the flow of a particle-laden thin film on an inclined plane, \textit{Phys. Fluids}, \textbf{21}, 083305 (2009).\\

\noindent 2. B. Cook,  O. Alexandrov and A. L. Bertozzi. Linear stability of particle-laden thin films, \textit{The European Physical Journal - Special Topics}, \textbf{166}, 1, 77-81 (2009).\\

\noindent 3. N. Murisic , J. Hob, V. Huc, P. Latterman, T. Koche, K. Linf, M. Mata, A.L. Bertozzi. Particle-laden viscous thin-film flows on an incline: Experiments compared with a theory based on shear-induced migration and particle settling, \textit{Physica D: Nonlinear Phenomena}
\textbf{240}, 20, 1661-1673 (2011). \\ \ \\ 

\noindent \textbf{Acknowledgements}\ \\ \

\noindent The authors would like to thank Dr. Sungyon Lee and Mr. Gilberto Urdaneta for their invaluable help and useful insight they
have provided in the laboratory and in the making of this video.

\end{document}